\documentclass[aps,prc,twocolumn,superscriptaddress,showpacs]{revtex4-1}

\usepackage[]{graphicx}
\usepackage[]{epsfig}
\usepackage{longtable}
\usepackage{hyperref}
\hypersetup{letterpaper, colorlinks=true, urlcolor=blue, citecolor=blue, linkcolor=blue, breaklinks=true}
\usepackage{breakurl}
\usepackage{amsthm}
\usepackage{amsmath}
\usepackage{booktabs}
\usepackage{subfigure}
\renewcommand{\vec}[1]{\mathbf{#1}}
\begin{document}

\title{Two-Neutron Sequential Decay of $^{24}$O}


\author{M.D.~Jones}
    \email[]{jonesm@nscl.msu.edu}
    \affiliation{National Superconducting Cyclotron Laboratory, Michigan State University, East Lansing, MI 48824, USA}
    \affiliation{Department of Physics and Astronomy, Michigan State University, East Lansing, MI 48824, USA}
\author{N.~Frank}
    \affiliation{Department of Physics and Astronomy, Augustana College, Rock Island, IL 61201, USA}
\author{T.~Baumann}
    \affiliation{National Superconducting Cyclotron Laboratory, Michigan State University, East Lansing, MI 48824, USA}
\author{J.~Brett}
    \affiliation{Department of Physics, Hope College, Holland, MI 49422-9000, USA}
\author{J.~Bullaro}
    \affiliation{Department of Physics and Astronomy, Augustana College, Rock Island, IL 61201, USA}
\author{P.A.~DeYoung}
    \affiliation{Department of Physics, Hope College, Holland, MI 49422-9000, USA}
\author{J.E.~Finck}
    \affiliation{Department of Physics, Central Michigan University, Mount Pleasant, MI 48859, USA}
\author{K.~Hammerton}
   \affiliation{National Superconducting Cyclotron Laboratory, Michigan State University, East Lansing, MI 48824, USA}
   \affiliation{Department of Chemistry, Michigan State University, East Lansing, MI 48824, USA}
\author{J.~Hinnefeld}
    \affiliation{Deparment of Physics and Astronomy, Indiana University South Bend, South Bend, IN 46634-7111, USA}
\author{Z.~Kohley}
   \affiliation{National Superconducting Cyclotron Laboratory, Michigan State University, East Lansing, MI 48824, USA}
   \affiliation{Department of Chemistry, Michigan State University, East Lansing, MI 48824, USA}
\author{A.N.~Kuchera}
    \affiliation{National Superconducting Cyclotron Laboratory, Michigan State University, East Lansing, MI 48824, USA}
\author{J.~Pereira}
   \affiliation{National Superconducting Cyclotron Laboratory, Michigan State University, East Lansing, MI 48824, USA}
\author{A.~Rabeh}
    \affiliation{Department of Physics and Astronomy, Augustana College, Rock Island, IL 61201, USA}
\author{W.F.~Rogers}
    \affiliation{Deparment of Physics, Westmont College, Santa Barbara, CA 93108, USA}
\author{J.K.~Smith}
    \altaffiliation[Present Address:]{ TRIUMF, 4004 Wesbrook Mall, Vancouver, British Columbia, V6T 2A3 Canada}
    \affiliation{National Superconducting Cyclotron Laboratory, Michigan State University, East Lansing, MI 48824, USA}
    \affiliation{Department of Physics and Astronomy, Michigan State University, East Lansing, MI 48824, USA}
\author{A.~Spyrou}
    \affiliation{National Superconducting Cyclotron Laboratory, Michigan State University, East Lansing, MI 48824, USA}
    \affiliation{Department of Physics and Astronomy, Michigan State University, East Lansing, MI 48824, USA}
    \author{S.L.~Stephenson}
    \affiliation{Department of Physics, Gettysburg College, Gettysburg, PA 17325, USA} 
\author{K.~Stiefel}
    \affiliation{National Superconducting Cyclotron Laboratory, Michigan State University, East Lansing, MI 48824, USA}
    \affiliation{Department of Chemistry, Michigan State University, East Lansing, MI 48824, USA}
\author{M.~Tuttle-Timm}
    \affiliation{Department of Physics and Astronomy, Augustana College, Rock Island, IL 61201, USA}
\author{R.G.T.~Zegers}
    \affiliation{National Superconducting Cyclotron Laboratory, Michigan State University, East Lansing, MI 48824, USA}
    \affiliation{Department of Physics and Astronomy, Michigan State University, East Lansing, MI 48824, USA}
    \affiliation{Joint Institute for Nuclear Astrophysics -- Center for the Evolution of the Elements, Michigan State University, East Lansing, Michigan 48824, USA}
\author{M.~Thoennessen}
    \affiliation{National Superconducting Cyclotron Laboratory, Michigan State University, East Lansing, MI 48824, USA}
    \affiliation{Department of Physics and Astronomy, Michigan State University, East Lansing, MI 48824, USA}

\date{\today}

\begin{abstract}
    
    A two-neutron unbound excited state of $^{24}$O was populated through a (d,d') reaction at 83.4 MeV/nucleon. A state at $E = 715 \pm 110$ (stat) $\pm 45 $ (sys) keV with a width of $\Gamma < 2$ MeV was observed above the two-neutron separation energy placing it at 7.65 $\pm$ 0.2 MeV with respect to the ground state. Three-body correlations for the decay of $^{24}$O $\rightarrow$ $^{22}$O + $2n$ show clear evidence for a sequential decay through an intermediate state in $^{23}$O. Neither a di-neutron nor phase-space model for the three-body breakup were able to describe these correlations.

\end{abstract}

\doi{10.1103/PhysRevC.92.051306}

\maketitle


Nuclei near or beyond the dripline offer a valuable testing ground for nuclear theory, because they can exhibit phenomena that otherwise might not be observed in more
stable nuclei \cite{RevModPhys.84.567, ThoennessenRev, ThomasRev}. For systems which decay by emission of two particles, the three-body correlations of two-proton and two-neutron unbound nuclei provide valuable insight into the features of their decay mechanisms. First predicted by Goldansky in 1960 \cite{Goldansky1960482}, two-proton radioactivity has been observed in $^{45}$Fe \cite{PhysRevLett.89.102501}, and genuine three-body decays along with their angular correlations have been observed in many two-proton unbound systems, e.g. $^{19}$Mg, $^{16}$Ne \cite{PhysRevC.77.061303} and $^{6}$Be \cite{PhysRevLett.109.202502}. 

Analogously, two-neutron unbound systems and their three-body correlations have been measured in several neutron-rich nuclei including $^{5}$H \cite{PhysrevC.72.064612}, $^{10}$He \cite{PhysRevLett.108.202502, Johansson201066}, $^{13}$Li \cite{Johansson201066, PhysRevC.87.011304}, $^{14}$Be \cite{PhysRevLett.111.242501}, $^{16}$Be \cite{PhysRevLett.108.102501} and  $^{26}$O \cite{PhysRevC.91.034323}, with the last one showing potential for two-neutron radioactivity
\cite{PhysRevLett.110.152501, PhysRevC.88.034313}.  Several of these systems have been interpreted to decay by emission of a di-neutron \cite{PhysRevLett.108.102501, PhysRevC.87.011304}. So far, there has been no correlation measurement of a state decaying by sequential emission of two neutrons, although evidence for sequential decay has been reported for the high-energy continuum of $^{14}$Be \cite{PhysRevLett.111.242501}.

The structure of $^{24}$O, the heaviest bound isotope for which the neutron drip-line is established \cite{Thoennessen201243}, has been well studied and there is substantial evidence for the appearance of a new magic number $N=16$ \cite{PhysRevLett.102.152501, PhysRevLett.109.022501, Hoffman200917, PhysRevC.69.034312, PhysRevLett.100.152502}. Two unbound resonances have been observed in $^{24}$O above the one-neutron separation energy  \cite{PhysRevLett.109.022501,
PhysRevC.83.031303,WarrenO24}, and there is also evidence for resonances above the two-neutron separation energy \cite{PhysRevLett.109.022501, PhysRevC.83.031303} around 7.5 MeV. The first tentative evidence that one of these resonances decays by sequential emission of two neutrons was deduced from a measurement of two discrete neutron energies in coincidence similar to a $\gamma$-ray cascade \cite{ PhysRevC.83.031303}. In this Rapid Communication we present the first observation of a two-neutron sequential decay, exposed by energy and angular correlations, in $^{24}$O.


The experiment was performed at the National Superconducting Cyclotron Laboratory (NSCL), where a 140 MeV/nucleon $^{48}$Ca beam impinged upon a $^{9}$Be target with a thickness of 1363 mg/cm$^{2}$ to produce an $^{24}$O beam at 83.4 MeV/nucleon with a purity of $30\%$ at the end of the A1900 fragment separator. The $^{24}$O fragments could be cleanly separated from the other contaminants by time-of-flight in the off-line analysis. The secondary beam then proceeded to the experimental area where it
impinged upon the Ursinus College Liquid Hydrogen Target, filled with liquid deuterium (LD$_{2}$), at a rate of approximately 30 particles per second. The LD$_{2}$ target is cylindrical with a diameter of 38 mm and a length of 30 mm, is sealed with Kapton foils 125 $\mu$m thick on each end, and is based on the design of Ryuto \emph{et al.} \cite{hydrogen_target}.  For the duration of the experiment, the target was held slightly above the triple point of
deuterium at $19.9 \pm 0.25$ K and $850 \pm 10$ Torr, and wrapped with 5$\mu$m of aluminized Mylar to ensure temperature stability. The Kapton windows of the target were deformed by the pressure differential between the target cell and the vacuum thus adding to the thickness of the target.

A (d,d') reaction excited the $^{24}$O beam above the two-neutron separation energy, $S_{2n}=6.93 \pm 0.12$ MeV \cite{AME2012}, which then promptly decayed. The resulting charged fragments were swept 43.3$^{\circ}$ from the beam axis by a 4-Tm superconducting sweeper magnet \cite{sweeper} into a series of position and energy-sensitive charged particle detectors. Two cathode-readout drift chambers (CRDCs), separated by 1.55 m, were placed after the sweeper and measured the position of the charged
fragments. Immediately following the CRDCs was an ion-chamber which provided a measurement of energy loss, and a thin (5 mm) $dE$ plastic scintillator which was used to trigger the system readout and measure the time-of-flight (TOF). Finally, an array of CsI(Na) crystals stopped the fragments and measured the remaining total energy. The position and momentum of the fragments  at the target were reconstructed using an inverse transformation matrix \cite{nathan_nim}, obtained from the program COSY INFINITY \cite{cosy}.

Element identification was accomplished via a $\Delta E$ vs. TOF measurement, and isotope identification of $^{22}$O was accomplished through correlations between the time-of-flight, dispersive angle, and position of the fragments. Additional details on this technique can be found in Ref. \cite{GregPRC2012}. The neutrons emitted in the decay of $^{24}$O traveled 8 m undisturbed by the magnetic field towards the Modular Neutron Array (MoNA) \cite{mona_nimA} and the Large-area
multi-Institutional Scintillator Array (LISA). MoNA and LISA each consist of 144  200 x 10 x 10 cm$^{3}$ bars of plastic scintillator with photomultiplier tubes on both ends which provide a measurement of the neutron time-of-flight and position. MoNA and LISA each contain nine vertical layers with 16 bars per layer. The combined array was configured into three blocks of detector bars. LISA was split into two tables 4 and 5 layers thick, with the 4 layer table placed at 0$^{\circ}$ in front of MoNA,
while the remaining portion of LISA was placed off-axis and centered at 22$^{\circ}$. The resulting angular coverage was from 0$^{\circ}$ $\le \theta \le$ 10$^{\circ}$  in the laboratory frame for the detectors placed at 0$^{\circ}$, and 15$^{\circ}$ $\le \theta \le$ 32$^{\circ}$ for the off-axis portion. Together MoNA, LISA, and the charged particle detectors provide a complete kinematic measurement of the neutrons and charged particles, from which the decay of $^{24}$O can be reconstructed.


The momentum vectors of the neutrons in coincidence with $^{22}$O were calculated from their locations in MoNA-LISA. Neutron interactions were separated from background $\gamma$ rays by requiring a threshold of 5 MeV of equivalent electron energy (MeVee) on the total-charge deposited. In addition, a time-of-flight gate on prompt neutrons was also applied. 
\begin{figure}
\epsfig{file=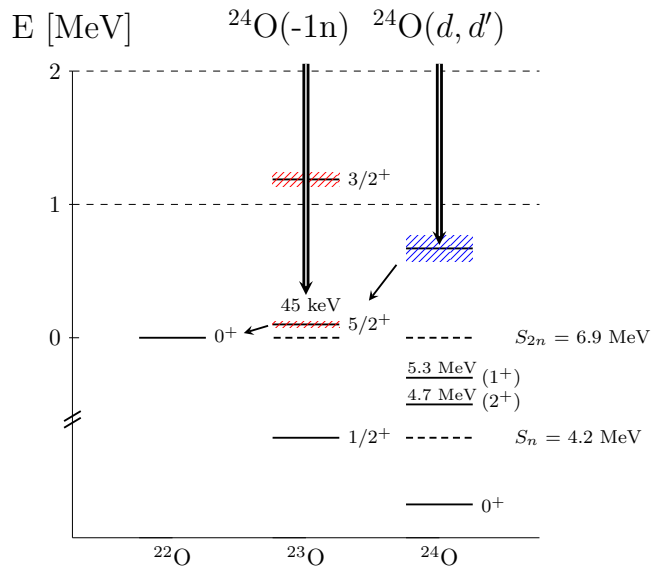, width=\columnwidth, angle=0}\\
\caption{(Color online) Level scheme for the population of unbound states in $^{23}$O and $^{24}$O from neutron-knockout and inelastic excitation. Hatched areas indicate approximate widths.}
\label{fig:scheme}
\end{figure}

The $N$-body decay energy is defined as $E_{\mathrm{decay}} = M_{\mathrm{Nbody}} - M_{^{22} \mathrm{O}} - \sum_{i=1}^{i=N-1}m_{\mathrm{n}}$,
where $M_{\mathrm{Nbody}}$ is the invariant mass of the $N$-body system, $M_{^{22}\mathrm{O}}$ the mass of $^{22}$O, and $m_{\mathrm n}$ the mass of a neutron. The invariant mass for an $N$-body system was calculated from the experimentally measured four-momenta of $^{22}$O and the first $N-1$ time-ordered interactions in MoNA-LISA. In this analysis, we consider both the two- and three-body decay energies.
The \textbf{T} and \textbf{Y} Jacobi coordinate systems were used to define the energy and angular correlations in the three-body decay of $^{24}$O. These correlations can be described by an energy distribution parameter $\epsilon = E_{x}/E_{T}$, and an angle $\cos(\theta_{k}) = \vec{k_{x}} \cdot \vec{k_{y}} / (k_{x}k_{y})$ between the Jacobi momenta $\vec{k_{x}}$ and $\vec{k_{y}}$  \cite{RevModPhys.84.567}. 

In constructing the three-body system and its correlations, it is crucial to identify events which are true 2n events as opposed to a single neutron scattering twice. This is accomplished by selection on the relative distance $D_{12}$ and relative velocity $V_{12}$ between the first two interactions in MoNA-LISA. By requiring a large relative distance $D_{12}$, events which scatter nearby are removed, and events with a clear spatial separation are selected. Since a neutron will lose energy when
it scatters, an additional cut on $V_{12}$ will also remove scattered neutrons. In this analysis, we require $D_{12} >$ 50 cm, and $V_{12}  >$ 12 cm/ns which is the beam velocity. This technique has been used in several measurements of two-neutron unbound nuclei to discriminate against 1n scatter \cite{PhysRevLett.109.232501, PhysRevLett.108.142503, PhysRevLett.108.102501, PhysRevC.87.011304, PhysRevC.83.031303, PhysRevC.90.024309, JennaLi11, PhysRevLett.96.252502,
Marques2000109}.  To further enhance the 2n signal, an additional threshold of 5 MeVee is applied to every hit recorded in MoNA-LISA. 

\begin{figure}
\epsfig{file=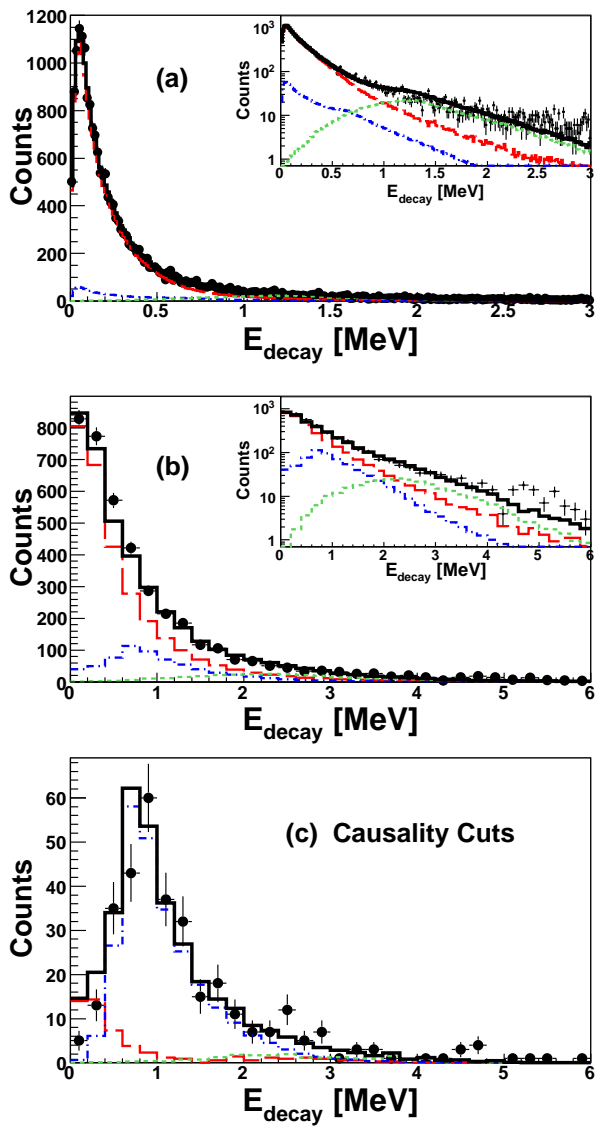, width=\columnwidth, angle=0}\\
\caption{(Color online) (a) 1$n$ decay energy spectra for $^{23}$O with contributions from neutron-knockout and inelastic excitation. (b) Three-body decay energy for $^{22}$O + $2n$ for all multiplicities $\ge 2$. 
    (c) Three-body decay energy with causality cuts applied. Direct population of the $5/2^{+}$ state and $3/2^{+}$ state in $^{23}$O are shown in dashed/red and dotted/green respectively. The 2$n$ component coming from the sequential decay of $^{24}$O is shown in dot-dash/blue and decays through the $5/2^{+}$ state.  The sum of all components is shown in black.} 
\label{fig:1n_mult}
\end{figure}

A Monte Carlo simulation was used to model the decay of $^{24}$O and included the beam characteristics, the reaction mechanism, and subsequent decay. The efficiency, acceptance, and resolution of the charged particle detectors following the dipole sweeper magnet and MoNA-LISA were fully incorporated into the simulation making the result directly comparable with experiment. The neutron interactions in MoNA-LISA were modeled with GEANT4 \cite{GEANT} and a modified version of MENATE$\_$R
\cite{MENATER} where the $^{12}C(n,np)^{11}B$ inelastic cross-section was modified to better agree with an earlier measurement \cite{PhysRev.90.224}. 

In principle, $^{22}$O can be populated by multiple paths in this experiment as illustrated in Fig. \ref{fig:scheme}. The first is by neutron-knockout to an unbound state in $^{23}$O; another is by inelastic excitation of the beam above the two-neutron separation energy. Hence, it is important to consider both the one and two-neutron decay energy spectra. This is done by a simultaneous minimization of the log-likelihood ratio on three experimental histograms: (a) the $^{22}$O + 1n decay
energy, (b) $^{22}$O + 2n decay energy, and (c) the $^{22}$O + 2n decay energy with the causality cuts shown in Fig. \ref{fig:1n_mult}. This method provides additional constraints over fitting each histogram individually.

To allow for the direct population of $^{23}$O by one-neutron knockout, we included the decay from two previously reported states in $^{23}$O: the low-lying sharp resonance at 45 keV  and the first-excited state at 1.3 MeV \cite{Frank2008199, PhysRevLett.99.112501, PhysRevLett.98.102502, PhysRevC.83.031303, Tshoo201419}. The two-neutron decay from $^{24}$O was modeled following the formalism of Volya \cite{volya_EPJ} and was treated as a single state in $^{24}$O which could decay sequentially through either of the two states in $^{23}$O. 

In this formalism, a distribution for the relative energy of the two neutrons $E_{r} = \epsilon_{1} - \epsilon_{2}$ is calculated as a function of the total decay energy $E$. The first neutron, with kinetic energy $\epsilon_{1}$, decays from an initial state with energy $E_{1}$ and width $\Gamma_{1}$, to an intermediate unbound state $E_{2}$, $\Gamma_{2}$, which proceeds to decay by emitting another neutron with kinetic energy $\epsilon_{2}$. Assuming a spin anti-symmetric pair of neutrons, the total amplitude for the decay becomes:

\begin{equation*}
    \begin{split}
        &A_{T}(\epsilon_{1}, \epsilon_{2})  \\
                                           &= \frac{1}{\sqrt{2}} \left( \frac{A_{1}(\epsilon_{1})A_{2}(\epsilon_{2})}{\epsilon_{2} - \left[ E_{2} - \frac{i}{2}\Gamma_{2}(\epsilon_{2})\right]} +  \frac{A_{1}(\epsilon_{2})A_{2}(\epsilon_{1})}{\epsilon_{1} - \left[E_{2} - \frac{i}{2}\Gamma_{2}(\epsilon_{1})\right]}     \right), 
                                     \end{split}                                                                      
\end{equation*}

Where $A_{1}$ and $A_{2}$ are the single-particle decay amplitudes, and $\Gamma_{i}$ are proportional to the energy-dependent single-particle decay widths by a spectroscopic factor. The Fermi golden rule then gives the partial decay width as:
\[ \frac{d\Gamma(E)}{d\epsilon_{1}d\epsilon_{2}} = 2\pi \delta(E - \epsilon_{1} - \epsilon_{2}) |A_{T}(\epsilon_{1}, \epsilon_{2})|^{2},  \]
And the cross-section is approximated as an energy-dependent Breit-Wigner with the differential written in terms of the relative energy:
\[ \frac{d\sigma}{dE_{r}} \propto \frac{1}{(E - E_{1})^{2} + \Gamma_{T}^{2}(E)/4} \frac{d \Gamma(E)}{d E_{r}}, \]
Where the total width $\Gamma_{T}(E)$ is obtained from:
\[ \Gamma_{T} = \int dE_{r} \frac{d \Gamma(E)}{dE_{r}}. \]
It should be noted, that this formalism assumes that the two neutrons come from the same orbital and are coupled to a $J^{\pi} = 0^{+}$.


The best fit for the decay of $^{23}$O and $^{24}$O is shown in Fig. \ref{fig:1n_mult}. Using previously reported values for states in $^{23}$O \cite{PhysRevLett.99.112501, PhysRevC.83.031303, Frank2008199, Tshoo201419}, we obtain good agreement with the data.   The two-neutron decay energy spectrum with the causality
cuts is shown in Fig. \ref{fig:1n_mult} (c). The best fit for the three-body decay gives an energy of $E$ = $715 \pm 110$ (stat) $\pm 45$ (sys) keV, and $\Gamma < 2$  MeV, which agrees with previous measurement \cite{PhysRevC.83.031303}. Only an upper limit can be put on the width, since the width is dominated by the experimental resolution. The best-fit is shown using the single-particle decay width of $\Gamma_{spdw} = 120$ keV. Using a value of $S_{2n} = 6.93 \pm 0.12$ MeV \cite{AME2012}, places the state at an excitation energy of 7.65 $\pm 0.2$  MeV. 
No branching through the $3/2^{+}$ state in $^{23}$O was necessary to fully describe the data.
\begin{figure}
\epsfig{file=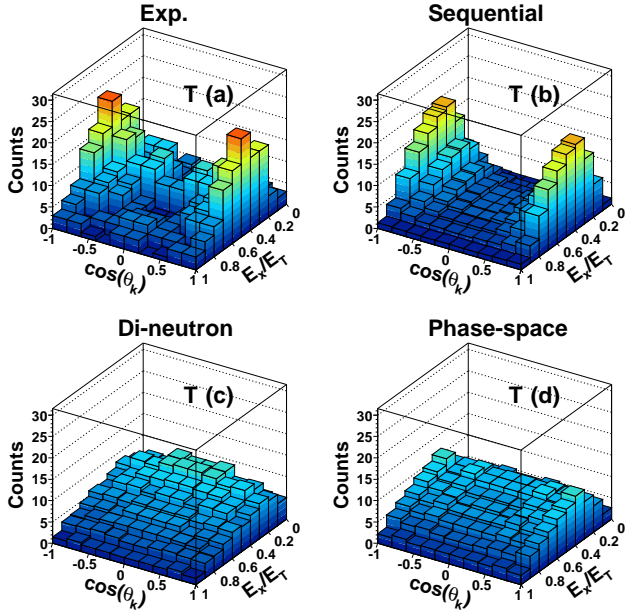, width=\columnwidth, angle=0}\\
\caption{(Color online) Jacobi relative energy and angle spectra in the \textbf{T} system for the decay of $^{24}$O $\rightarrow$ $^{22}$O + $2n$ with the causality cuts applied and the requirement that $E_{decay} < 4$ MeV. Shown for comparison are simulations of several three-body decay modes. A sequential decay (b), a di-neutron decay with $a=-18.7$ fm (c), and (d) a phase-space decay. The amplitudes are set by twice the integral of the three-body spectrum with causality cuts.}
\label{fig:comparison}
\end{figure}

\begin{figure}
\epsfig{file=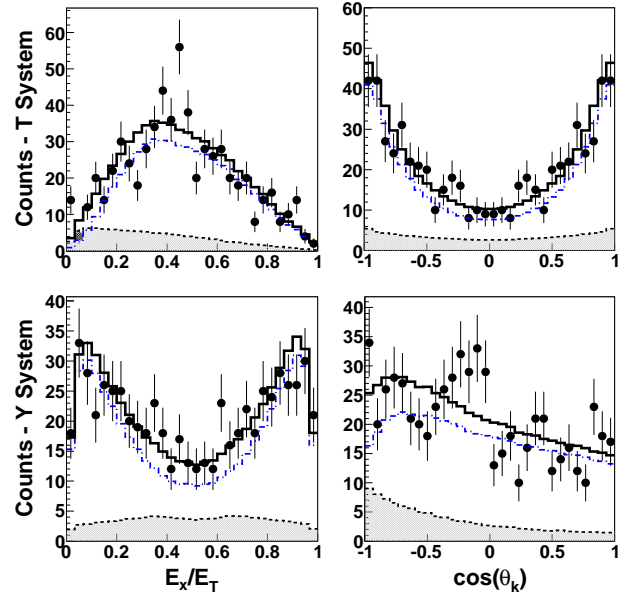, width=\columnwidth, angle=0}\\
\caption{(Color online) Jacobi relative energy and angle spectra in the \textbf{T} and \textbf{Y} systems for the decay of $^{24}$O $\rightarrow$ $^{22}$O + 2n with the causality cuts applied and the requirement that $E_{decay} < 4$ MeV. In dashed-blue is the sequential decay through the $5/2^{+}$ state in $^{23}$O. The remaining false 2$n$ components from the 1n decay of $^{23}$O are shown in shaded-grey. The sum of both components is shown in solid-black}
\label{fig:jacobi}
\end{figure}

The two- and three-body decay energies, along with the causality cuts, are well described by the sequential decay, where each decay proceeds by emission of an $L=2$ neutron. 
The data are largely dominated by the direct population of $^{23}$O, however the one-neutron decay is unable to describe the three-body energy with causality cuts and the corresponding three-body correlations.

The relative energy and angle in the \textbf{T} system is shown for three-body events with causality cuts in Fig. \ref{fig:comparison}(a).  In addition to the sequential decay (b), two other models were tested: (c) the di-neutron decay \cite{PhysRevC.74.064314}, and (d) an (uncorrelated) phase-space decay \cite{CERN1968}. In the di-neutron decay the two-neutrons are emitted as a pair which subsequently breaks up, and so they have an angular correlation that peaks at $-1$ in the
\textbf{Y}-system $\cos(\theta_{k})$. In contrast to the sequential and di-neutron emission, the phase-space decay assumes no correlations between the neutrons and distributes their energy evenly. 

It is evident that the two-neutron decay does not proceed by di-neutron nor by uncorrelated emission and is much better described by the sequential decay, demonstrating that the two-neutron decay passes through an intermediate state instead of directly populating $^{22}$O.


Fig. \ref{fig:jacobi} shows the best-fit of the energy and angle correlation in the \textbf{T} and \textbf{Y} system for a sequential decay. It includes contributions from false 2$n$ events shown in shaded gray. They are clearly distinct from the correlations for the sequential decay. Most notable is the $cos(\theta_{k})$ in the \textbf{T} system which exhibits strong peaks at $1$ and $-1$ with a valley in between. Similarly, the relative energy spectrum in the \textbf{Y}-system is
peaked around $0$ and $1$. The ratio $E_{x}/E_{T}$ in the \textbf{Y}-system is indicative of how the energy is shared between the neutrons, where a peak at $1/2$ implies equal sharing. The data show unequal sharing which indicates a sequential decay through a narrow state that is closer to the final state than it is to the initial state (or vice versa). 

This is what we expect given that the three-body state in $^{24}$O is at 715 keV, while the intermediate state in $^{23}$O is narrow and low-lying at 45 keV. In the two-proton decay of $^{6}$Be \cite{PhysRevLett.109.202502}, it was observed that the sequential decay was suppressed until the total decay energy was greater than twice the intermediate
state plus its width. Here, this condition is certainly fulfilled. The depth of the valley in the $E_{x}/E_{T}$ spectrum is slightly softened by contamination from false 2n events. 
In the analysis of the two-proton decay of $^6$Be the data could not be described by either of three simple models (di-proton, sequential, direct) and a more complex fully three-body dynamical calculation was used to describe the data. Although the present data show clear evidence for a sequential two-neutron decay in $^{24}$O, a similar three-body calculation would be valuable to fully understand the decay mechanism.

In summary, a state above the two-neutron separation energy in $^{24}$O was populated by inelastic excitation on a deuterium target. The data are well described by a single resonance at $E = 715 \pm 110$ (stat) $\pm$ 45 (sys) keV. Examination of the three-body Jacobi coordinates shows strong evidence for a sequential decay through a low-lying intermediate state in $^{23}$O at 45 keV. The di-neutron or phase-space models are unable to reproduce these correlations. Unlike other
systems that decay by emission of two-neutrons and show evidence for di-neutron decay \cite{PhysRevLett.108.102501, PhysRevC.87.011304}, the decay of $^{24}$O $\rightarrow$ $^{22}$O + 2n has unambiguously been determined to be a sequential process.  
\begin{acknowledgments}

We would like to thank L.A. Riley for the use of the Ursinus College Liquid Hydrogen Target. This work was supported by the National Science Foundation under Grants No. PHY14-04236, No. PHY12-05537, No. PHY11-02511, No. PHY11-01745, No. PHY14-30152, No. PHY09-69058 and No. PHY12-05357. The material is also based upon work supported by the Department of Energy National Nuclear Security Administration under Award No. DE-NA0000979.
\end{acknowledgments}
\bibliography{library.bib}
\end{document}